# Super-resolution Ultrasound Localization Microscopy through Deep Learning

Ruud. J.G. van Sloun[1]*, Oren Solomon[3], Matthew Bruce[4], Zin Z. Khaing[5], Hessel Wijkstra[1,2], Yonina C. Eldar[3], Massimo Mischi[1]

[1]Dept. of Electrical Engineering, Eindhoven University of Technology, Eindhoven, The Netherlands.
[2]Dept. of Urology, Academic Medical Center, University of Amsterdam, Amsterdam, The Netherlands.
[3]Dept. of Electrical Engineering, Techion – Israel Institute of Technology, Haifa, Israel.
[4]Applied Physics Laboratory, The University of Washington, Seattle, WA, USA.
[5]Dept. of Neurological Surgery, The University of Washington, Seattle, WA, USA.
*e-mail: r.j.g.v.sloun@tue.nl

## ABSTRACT

**Ultrasound localization microscopy has enabled super-resolution vascular imaging through precise localization of individual ultrasound contrast agents (microbubbles) across numerous imaging frames. However, analysis of high-density regions with significant overlaps among the microbubble point spread responses yields high localization errors, constraining the technique to low-concentration conditions. As such, long acquisition times are required to sufficiently cover the vascular bed. In this work, we present a fast and precise method for obtaining super-resolution vascular images from high-density contrast-enhanced ultrasound imaging data. This method, which we term Deep Ultrasound Localization Microscopy (Deep-ULM), exploits modern deep learning strategies and employs a convolutional neural network to perform localization microscopy in dense scenarios. This end-to-end fully convolutional neural network architecture is trained effectively using on-line synthesized data, enabling robust inference *in-vivo* under a wide variety of imaging conditions. We show that deep learning attains super-resolution with challenging contrast-agent densities, both *in-silico* as well as *in-vivo*. Deep-ULM is suitable for real-time applications, resolving about 70 high-resolution patches (128x128 pixels) per second on a standard PC. Exploiting GPU computation, this number increases to 1250 patches per second.**

## INTRODUCTION

Robust, precise, fast and cost-effective *in-vivo* microvascular imaging is a cornerstone for clinical management of diseases that are hallmarked by impaired or remodeled microvasculature, such as angiogenesis in cancer[1]. Contrast-enhanced ultrasound is a cost-effective modality, which combines ultrasound imaging with enhancement of blood through the use of ultrasound contrast agents, inert gas microbubbles that are sized similar to red blood cells[2]. Nevertheless, the spatial resolution of conventional contrast-enhanced ultrasound imaging is bound by the diffraction limit of sound. Being primarily determined by the adopted wavelength, this limit in practice manifests itself as an inherent trade-off between resolution and penetration depth, since acoustic waves suffer from increasing amounts of absorption at higher frequencies.

Recently, this trade-off was circumvented through the introduction of Ultrasound Localization Microscopy (ULM), where Nobel-prize-winning super-resolution concepts from optics (e.g. Photoactivation Localization Microscopy - PALM) are exploited and translated into the ultrasound imaging domain to achieve sub-wavelength resolution images of the vasculature[3,4]. By pinpointing individual microbubbles from diffraction-limited ultrasound data across a large sequence of imaging frames with sparse microbubble populations, i.e. low contrast-concentration, and combining all these position estimates into one frame, a super-resolved image is produced[5]. Errico *et al.* implemented this concept by acquiring over 75,000 frames of a fixed rat brain using an ultrafast ultrasound imaging scheme across 2.5 minutes[6]. While attaining such motion-free acquisition across this time span is feasible in confined laboratory environments, it is impossible in most clinical situations where the impact of motion can be severe.

Beyond mapping the accumulated microbubble localizations, one can also track detections across multiple frames[7,8]. This not only permits rendering of velocities and interpolation of the detections, but also removal of spurious clutter localizations. While nearest-neighbor data association schemes are common[6,7,9,10], Ackermann *et al.* showed that more robust assimilation can be obtained through dedicated motion models in a modified Markov chain Monte Carlo framework[11]. The latter was recently evaluated in a clinical setting[12], advocating the value of ULM for diagnostics in the context of vascular phenotyping of tumors. At the same time, the authors underline a major hurdle to overcome if ULM is to be broadly implemented in clinical practice: long measurements (even of about one minute) suffer from significant tissue motion. While mild in-plane motion can be compensated with sub-wavelength accuracy[13], registration errors for relatively large movements can be much larger than the localization precision of ULM. Moreover, out-of-plane components cannot be corrected in 2D.

ULM avoids the trade-off between resolution and penetration depth, but it gives rise to a new trade-off that balances localization precision, microbubble concentration and acquisition time. High image fidelity is attained when large amounts of bubbles are localized with high precision, posing a lower bound on the acquisition time of ULM. This bound can be relaxed significantly when high concentrations are used, with many high-precision localizations per frame. Moreover, the probability of actually filling all arterioles with microbubbles in a certain timespan increases with higher concentrations. Obtaining the required localization precision in data with such a dense population of microbubbles with overlapping signals is a challenging task however, yielding a scenario in which single-bubble localization algorithms break down. As such, standard ULM methods adhering to this microbubble-sparsity constraint still require long acquisition times, impairing broad translation in a clinical setting, where high contrast concentrations, limited time, significant organ motion and lower frame-rate imaging are common.

Algorithms based on sparse recovery have been developed specifically to cope with the overlapping point spread functions (PSFs) of multiple microbubbles[14–16]. These strategies pose the localization task as a sparse image recovery problem[14], in which bubbles with overlapping PSFs but distinct sparse locations on a dense grid can be resolved. Sparsity-based ultrasound super-resolution hemodynamic imaging (SUSHI)[15,17] expands upon this by considering the inherent temporal structure of the data, and performs sparse recovery in the temporal correlation domain. While successful localization of densely-spaced emitters has been demonstrated, even highly optimized fast recovery techniques involve a time-consuming iterative procedure. Fourier-domain fast iterative shrinkage-thresholding[15] improves dramatically over an image domain formulation, but computational time grows significantly with the field of view. In addition, the optimal settings of sparse-recovery methods can vary across frames due to e.g. time-varying microbubble densities.

Here we propose Deep-ULM, an ultrasound localization microscopy strategy based on deep learning[18], designed and trained to cope with high-concentration contrast-enhanced ultrasound (CEUS) acquisitions. We harness a fully convolutional neural network for super-resolution image reconstruction from dense images containing many overlapping microbubble signals, and show that the method is robust to varying imaging conditions and microbubble concentrations. The output of the network is not a set of position vectors but rather a high-resolution image in which the pixel values reflect recovered backscatter intensities. Our approach shares similarities with a recently introduced deep learning technique for single molecule fluorescence microscopy[19], albeit in a completely different field and setting. Image recovery using Deep-ULM is fast, and can be applied to any CEUS acquisition in which the PSF can be estimated, requiring minimal user expertise and no manual tweaking. We show that our approach outperforms both standard ULM as well as sparse recovery methods for high densities.

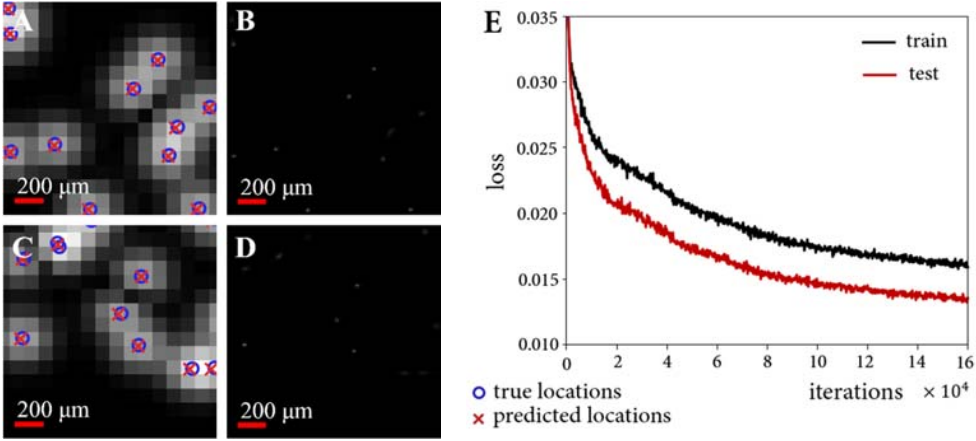

Figure 1. Deep-ULM for synthetic datasets. (**A**, **C**) Examples of synthetic datasets with different microbubble densities, generated using a point-spread-function model estimated from clinically acquired ultrasound data. (**B**, **D**) Corresponding Deep-ULM recoveries on a 15-$\mu$m spaced grid. The true locations of the microbubbles are marked as blue circles, and Deep-ULM predictions (on a discrete grid) as red crosses. (**E**) Train (with dropout) and test (disabled dropout) loss as a function of the number of iterations. The test loss is lower than the training loss so that the network generalizes well.

## RESULTS

**Training Deep-ULM on synthetic datasets**

Deep learning typically relies on the exploitation of large, representative datasets that enable the training of a robust network that generalizes well when employed in practice. While measuring sufficiently diverse CEUS inputs along with their super-resolved outputs is not trivial, the generation of realistic synthetic training data is in fact rather simple. To this end, we sample the real system PSF from CEUS images using a tool that enables manual selection of a few individual microbubbles across a few frames. We then automatically fit a rotated anisotropic Gaussian PSF model to the data to extract the PSF parameters. The generation of new synthetic data for training is straightforward (see Methods for details); each corresponding low-resolution CEUS input and super-resolved target represents the basis for a diverse training dataset involving a number of variations: randomly selected density that ranges from 0-260 microbubbles per $cm^2$, randomized microbubble locations along with backscatter intensities, white as well as colored background noise corrupting the CEUS images, and variance in the PSF parameters to account for uncertainty in their estimates. By introducing all these factors of variation, we strive to form a training set which is sufficiently complete and representative of real CEUS acquisitions.

A computational model should then be able to learn representations from this data through a hierarchy of nonlinear operations, having the capacity to perform an end-to-end mapping from diffraction-limited CEUS to super-resolved images. For this purpose, we adopted a fully convolutional network architecture based on an encoder-decoder structure, similar to the widely used U-net[20]. The encoder is trained to optimally convert the ultrasound image space into a feature space that contains all relevant microbubble position information, through convolutions and down-sampling operations. The decoder is trained to transform this feature space into a high-resolution, super-resolved frame via up-sampling and transposed convolutions.

Using backpropagation[18], we train the network to minimize the mean-squared-error between the super-resolved predictions and the ground-truth frames in batches of 256 synthetic imaging frames, across 20,000 iterations. As a unique batch of data is generated on-line for each iteration, the model's robustness and capacity to generalize to new cases is drastically improved. The latter is further supported by applying dropout during the training phase, by randomly disabling features at the encoded latent space with a probability of 0.5.

Figure 1 shows several examples of Deep-ULM applied to such synthetic datasets, with the reconstruction being on an 8 times up-sampled grid. Recovery of a high-resolution 128x128 patch using Deep-ULM takes less than 0.8 milliseconds on a GPU-equipped workstation, and about 14 milliseconds on a standard PC. The training and testing loss (a measure of resemblance between the network predictions and ground truth) monotonically decrease as a function of the number of iterations, showing no sign of overfitting. This can be attributed to the on-line synthetic data generation and dropout-based regularization. During testing, dropout is disabled, further pushing the loss down as a consequence of effective model ensemble averaging[21].

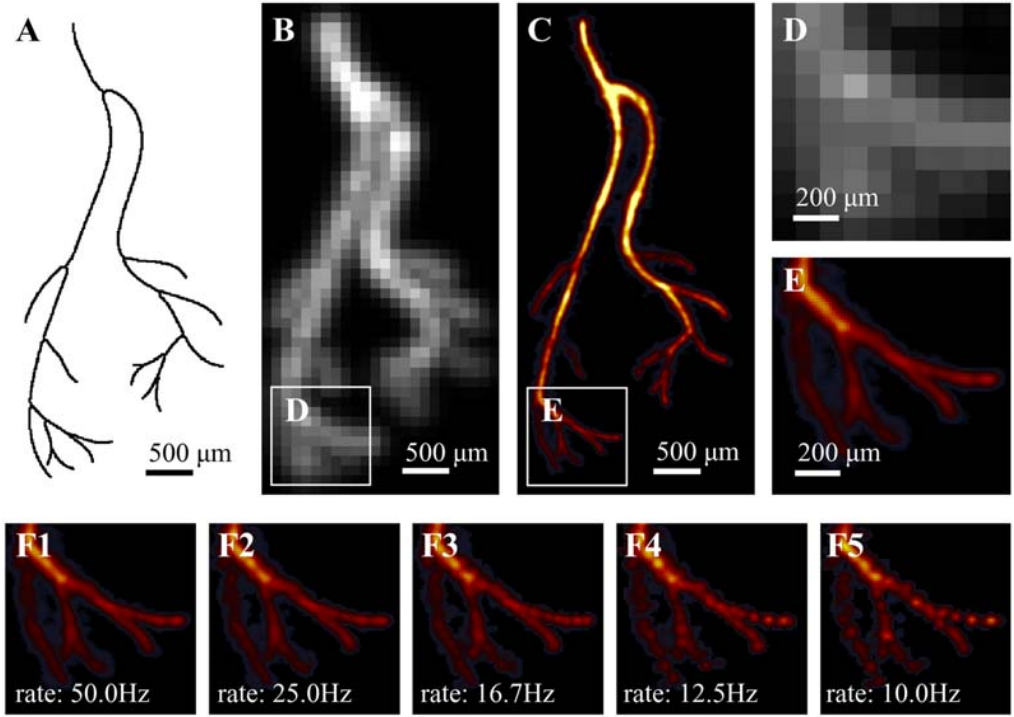

Figure 2. Deep-ULM on *in-silico* flow data compared to diffraction limited imaging. (**A**) Simulated vascular skeleton, (**B**) diffraction-limited maximum intensity persistence image, (**C**) Deep-ULM super-resolution reconstruction and (**D**,**E**) zooms of (**B**,**C**). (**F1-F5**) Deep-ULM reconstruction across 12 seconds with decreasing frame rates, displaying how dense localization on high-concentration simulations maintains reasonable fidelity even when very limited imaging frames are available. The actual physiological requirement is that vessels are sufficiently filled by the agent within the imaging time, which is relaxed by the use of high concentrations.

**Deep-ULM on *in-silico* flow of microbubbles through a branching vessel**

We first tested Deep-ULM *in-silico*, on a simulated CEUS acquisition of microbubble flow through a realistic bifurcating vessel. An infusion of microbubbles through this vascular phantom was mimicked by propagating microbubbles along the centerlines of the vessels with a velocity of 1 mm/s, along with an additional random component to introduce small velocity deviations among bubbles. The generation of microbubbles at the injection point was stochastic and followed a uniform distribution over time, at an average rate of ~12 microbubbles/s. This particular generation led to a mean distance of 79 $\mu$m between adjacent microbubbles at the first bifurcation, with a standard deviation of 75 $\mu$m. In the area around the final bifurcation these values were 229 $\mu$m and 165 $\mu$m, respectively. We then simulated the ultrasound imaging procedure by first convolving all bubbles with the modelled scanner PSF (an anisotropic 2D Gaussian modulated by the carrier frequency along fast time; wavelength of 200 $\mu$m). The simulated CEUS image is then formed by performing envelope detection (demodulation) on the attained radiofrequency lines and downsampling to a pixel spacing of 0.12 mm in both dimensions. We generated data for 12 seconds at a frame rate of 100 Hz. Deep-ULM detects over 20000 microbubbles across the generated 1200 frames, finely delineating the vascular architecture as shown in fig. 2. Because the method detects many microbubbles per frame, reconstructions at lower frame rates and shorter timespans become feasible. The impact of using such a reduced amount of imaging frames is evaluated in fig. 2F, showing that reconstructions with as few as 120 frames already display good fidelity.

In supplementary figs. 1 and 2, we show that when dealing with high densities, deep-ULM indeed outperforms a dedicated and optimized ultrasound localization microscopy approach that is based on computation of image centroids after deconvolution[6]; deep-ULM detects more microbubbles with a higher precision.

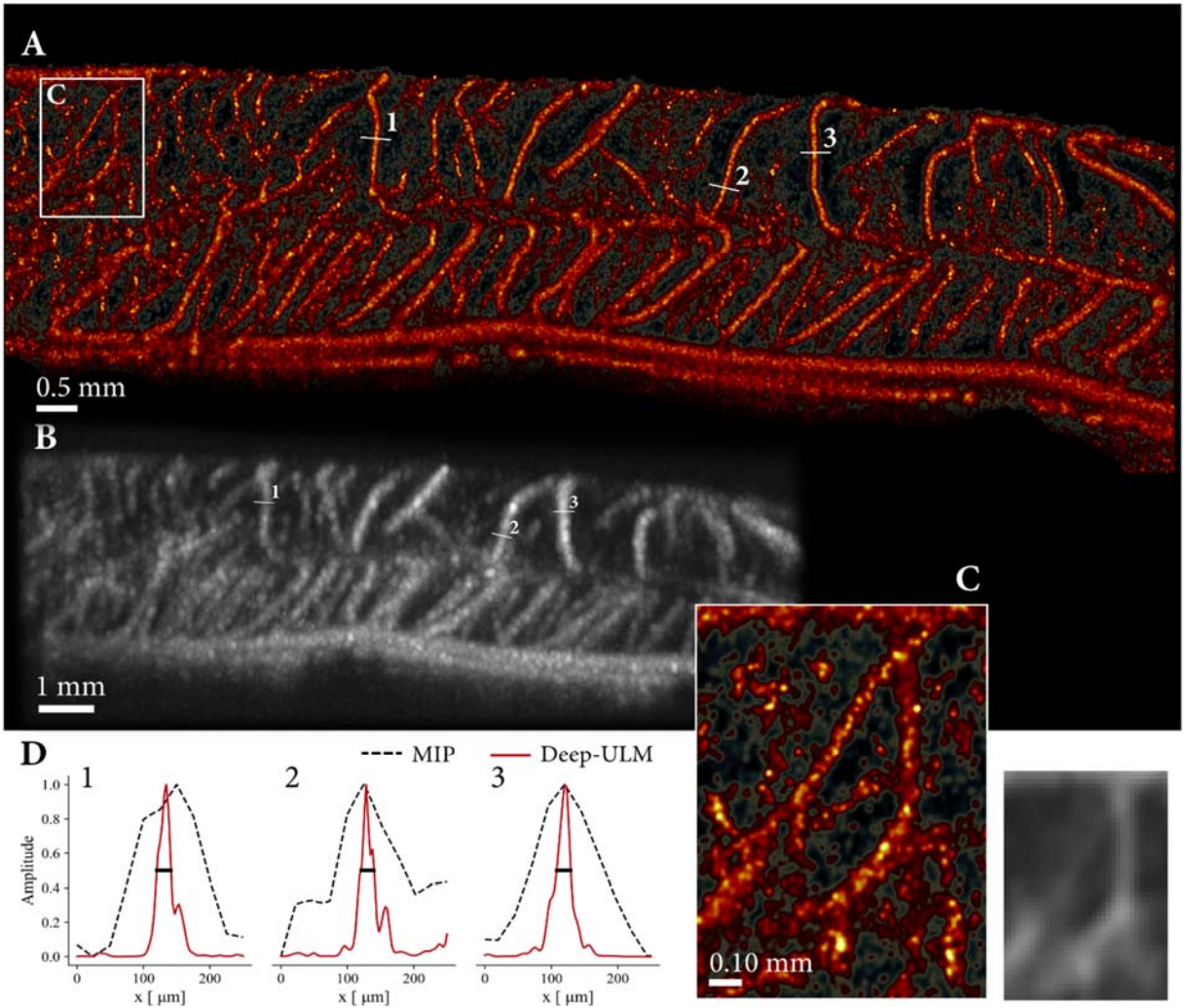

Figure 3. *In-vivo* Deep-ULM in a rat spinal cord. (**A**) Deep-ULM across 8-seconds acquired at a frame rate of 400 Hz, along with a (**B**) Maximum intensity projection image and (**C**) zoomed region of interest for Deep-ULM and the corresponding mean intensity image. (**D**) Intensity profiles of vessels, with their full-width-half maxima indicated by black horizontal lines, being 21 $\mu$m, 19 $\mu$m, and 20 $\mu$m for profiles 1, 2, and 3, respectively.

In these circumstances, deep-ULM also outperforms sparse-recovery based ULM[14], which in turn outperforms centroid-based ULM in terms of detection rate and localization precision (supplementary fig. 1). Deep-ULM is moreover about 4 orders of magnitude faster than the time-consuming sparse-recovery procedure. An additional comparison between frame-by-frame sparse-recovery, SUSHI[15], and Deep-ULM on several sub-diffraction spaced parallel streams shows the multi-frame model used in SUSHI to outperform standard frame-by-frame sparse-recovery, but Deep-ULM displays again the best performance (supplementary fig. 2).

### *In-vivo* Deep-ULM of a rodent spinal cord with high-frame-rate CEUS

We proceed to apply Deep-ULM *in-vivo*, using a high-frame-rate (400 Hz) CEUS scan of a rat spinal cord acquired with a Verasonics Vantage ultrasound research scanner[22]. We retrained the neural network based on an estimate of the PSF parameters of this system (obtained using the tool described earlier), and performed Deep-ULM on an 8-second acquisition to obtain a super resolved image. Recovery on the 8 times up-sampled (4096x1328) grid took ~100 milliseconds per complete imaging frame. In total, over 600,000 localizations were attained. Figure 3 shows how the method achieves super-resolution image recovery, resolving vessels beyond the diffraction limit. A spatial resolution of about 20-30 µm was achieved, estimated by measuring the full-width-half-maxima of several profiles of arterioles that carried a sufficiently large amount of microbubbles (see fig. 3D). This was a 4-5 fold improvement with respect to the image resolution of the maximum intensity projection image for these profiles.

In the supplementary materials (supplementary fig. 5), we also show initial feasibility of Deep-ULM in a clinical setting, with a standard clinical ultrasound system, contrast-agent dose, and protocol.

## DISCUSSION

Ultrasound localization microscopy (ULM) has enabled researchers to achieve extraordinary and unprecedented resolution in vascular ultrasound, no longer hindered by the diffraction limit of sound. Yet, its harsh limitations in terms of allowable contrast-agent concentrations lead to long acquisition times, and have spurred research in the direction of solving the high-density problem. Although recent methods exploiting sparse-recovery strategies do indeed allow for higher concentrations[14,15], they come at a high computational cost. In this paper, we show how deep learning enables a machine to learn how to perform efficient ULM in challenging high-density scenario's, requiring nothing more than an estimate of the local PSF of the image system. Notably, the network architecture, settings, and training procedure remained unchanged across the widely different *in-silico* and *in-vivo* experiments; the method was simply used "as is".

Deep-ULM uses a convolutional neural network that is trained using synthetic datasets that consist of ground truth microbubble backscatter amplitudes on a fine grid along with their corresponding CEUS ultrasound images. The method's performance depends on the capacity of the network to learn how to solve this sparse-recovery problem in an efficient manner, by learning a nonlinear function that maps low-resolution B-mode images to super-resolved localizations. On the other hand the quality and representability of the synthetic data for the actual acquisitions used during inference plays a major role. To improve robustness with respect to the latter, uncertainty in the estimated ultrasound scanner parameters is incorporated by introducing a variance in the adopted PSF model parameters across the dataset.

The neural network was designed based on an encoder-decoder principle to perform the end-to-end mapping between the input images and their targets; an architectural approach that has been widely adopted for various segmentation and image enhancement problems[20,23,24]. The total number of convolutional layers in our deep net amounts to 15, which yielded sufficient capacity to perform the desired sparse-recovery functionality, while not overfitting. The latter thrives with our on-line training data generation and the use of a relatively thin bottleneck latent layer with 50% dropout, effectively exploiting an ensemble of trained encoder models at the inference stage. With this deep network, super-resolution recovery of low-resolution images takes about 14 milliseconds per patch on a regular PC, and even less than 0.8 milliseconds when exploiting GPU computation. One could push this number further down by using model compression techniques[25], such as learning less complex models to replicate the current model's functionality through knowledge distillation[26].

Being trained to deal with concentrations as high as 260 microbubbles per $cm^2$, our experiments show that Deep-ULM indeed performs well for high densities; conditions in which single particle localization algorithms based on image centroids break down. While sparsity-driven algorithms[14,15] improve upon centroid-based localization, Deep-ULM outperforms them in both localization precision and speed, being about 4 orders of magnitude faster. Nevertheless, also for Deep-ULM, higher densities pose greater challenges for the algorithm. Although the maximum admittable concentration given a desired precision is significantly boosted, it will inherently depend on the signal to noise ratio of the ultrasound acquisition.

The ability to handle such high microbubble concentrations has significant implications for translation into clinical applications. Alleviating the very demanding temporal constraints of standard ULM by faster coverage of the relevant arterioles is a necessity rather than a luxury in many diagnostic settings, where time is scarce and the impact of organ movement across the acquisition becomes significant. With ultrafast high-frame-rate ultrasound imaging architectures finding their way into clinical scanners, a super-resolution method requiring less than 1000 frames can achieve sub-second temporal resolution, thereby drastically improving real-time clinical utility while at the same time mitigating those severe motion artifacts[14,15].

Opacic *et al.*[12] showed that application of super-resolution ultrasound in a clinical setting is feasible when using dedicated motion compensation and frame clustering strategies. While this holds great promise, a significant amount of acquisitions were excluded due to severe motion artifacts that could not be compensated for. Combining the above methodology with methods that exploit higher densities to reduce the acquisition time, such as Deep-ULM, could potentially bridge this practical gap.

While the present method is implemented for 2D imaging, the ability to perform 3D ULM in a fast and data-efficient manner would be a cornerstone for many of its purposes. Operating in a low-concentration regime, traditional ULM would require acquisition, transfer and storage of an enormous amount of volumes, precluding its current use[27]. On the other hand, Deep-ULM efficiently deals with higher-concentrations, significantly lowering the required amount of acquisitions. Its future translation into 3D might therefore actually be possible.

In this work, Deep-ULM was trained to localize individual microbubbles, without incorporating any structural priors on the vascular architecture or microbubble dynamics. Including such priors in the model has the potential to further improve image fidelity, and is part of future work. We note that care should be taken to ensure that such models generalize well to pathological conditions by choosing appropriate priors, or exploiting training data that also represents these diseased cases.

Deep-ULM enables high-fidelity super-resolution vascular ultrasound imaging under challenging conditions. It operates at a high very recovery speed and does not require manual tweaking by an expert user, opening vast new possibilities for localization microscopy in ultrasound imaging.

## MATERIALS AND METHODS

*In-silico* **microbubble flow and imaging simulation**

Flow of microbubbles through an artificial vascular network was simulated by propagating particles along streamlines with a specific velocity, comprising a deterministic part, as well as a multiplicative random component, i.e.: $\bar{v}(x, y, t) = \max(0, \bar{v}_{det}(x, y, t) \cdot \mathcal{N}(\mu = 1, \sigma = 1))$. 140 particles were infused at the injection point by randomly drawing particle injection times from a uniform distribution across a 12-second timespan, leading to the generation of approximately 12 particles per second. Ultrasound imaging of this process was simulated by modelling the scanner's point spread function as a bivariate Gaussian, modulated by the ultrasound wave frequency. The standard deviation in the axial and lateral direction were set to 0.14 and 0.16 mm respectively. The frequency was set to 7 MHz, approximating the response of a nonlinearly resonating microbubble to an ultrasound transmit frequency of 3.5 MHz after fundamental mode suppression (e.g. bandpass filtering or pulse inversion). The image was formed by demodulating the radiofrequency scan lines originating from the summed contributions of all microbubble responses trough the Hilbert transform. Frames were constructed at a rate of 100 Hz, and the pixel dimensions were 0.12 x 0.12 mm.

*In-vivo* **rat spinal cord contrast-enhanced ultrasound acquisitions and pre-processing**

The animal experiments were performed at the University of Washington, Seattle, WA, USA, with prior approval from the University of Washington's Institutional Animal Care and Use Committee (IACUC). All appropriate guidelines from the University's Animal Welfare Assurance (A3464-01) as well as the NIH Office of Laboratory Animal Welfare (OLAW) were followed. A 250-grams female Sprague Dawley rat (Harlan Labs, Indianapolis, IN) was anesthetized using isoflurane (5 % to induce and 2.5 – 3 % to maintain), and the area overlying the T7/T8 vertebrae was shaved, cleaned and sterilized. After dissection of paraspinal muscles, a laminectomy was performed to expose the spinal cord from T6 to T10. High frame rate CEUS acquisitions of the cord were performed with a Vantage ultrasound research platform (Verasonics, Seattle, WA, USA), using a linear array transducer (Vermon, Tours, France). The transmit center frequency was 15 MHz with 90% bandwidth in receive. An intravenous injection of 0.15-mL Definity® (Lantheus, New Jersey, USA) contrast agent followed by a 0.2-mL saline flush was administered via the tail vein using a catheter (BF-27-01, SAI Infusion Technologies, Lake Villa, IL, USA). We then waited about 3 minutes for the concentration to drop. A 5-angle plane wave amplitude modulated sequence was adopted[28], using delay-and-sum beamforming in receive. The plane wave frame rate was 30 kHz to avoid motion artifacts when compounding the 5 angles. Compounded images were obtained at a rate of 400 Hz. The IQ data were then wall filtered (Butterworth high-pass of order 20 with a cutoff at 50Hz) to suppress tissue clutter and enhance the response to microbubbles, and subsequently envelope detected through the Hilbert transform.

*In-vivo* **human prostate contrast-enhanced ultrasound acquisitions and pre-processing**

The *in-vivo* CEUS investigation was performed at the AMC University Hospital (Amsterdam, The Netherlands). The study received prior approval from the local institutional review board, and patients signed informed consent (study registered under ClinicalTrials.gov identifier NCT01481441). The passage of a microbubble bolus through a human prostate was obtained using an intravenous injection of 2.4-mL SonoVue® (Bracco, Milan, Italy), and consecutively imaged using a 2D transrectal ultrasound probe (C10-3v) and a Philips iU22 ultrasound system (Philips Heathcare, Bothell, WA). A contrast-specific imaging mode based on a power modulation pulse scheme at 3.5 MHz was used to enhance sensitivity to microbubbles while suppressing linear backscattering from tissue. The pixel dimensions are 0.15 x 0.15 mm and the frame rate was 10 Hz. The acquisition was performed during 120 seconds, recording the full in- and out-flow of the bolus. We selected a 30-second window during the washout phase to maintain a relatively stable concentration for our analysis. The measured grey-levels were then converted to acoustic intensities through a lookup table describing the ultrasound scanner's compression function.

**Motion compensation**

For motion compensation, we first extracted the tissue signal from the data (raw IQ and fundamental mode intensity images for the rat spine and human prostate CEUS acquisitions, respectively) by performing a singular value decomposition on the space-time data (i.e. a matrix of which the columns are vectorized frames), and attributing the first few singular values (describing components with high spatiotemporal coherence) to tissue[14]. We then computed the required rigid transformations that map each resulting frame to the first frame in the loop. The maximum measured motion was about 75 $\mu$m for the rat spinal cord sequence, and 200 $\mu$m for the clinical prostate scans.

**Synthetic training data generation**
We generated target patches containing multiple microbubbles with various intensities on a high-resolution grid. A broad spectrum of contrast-agent-concentrations was simulated, randomly drawn from a uniform distribution between 0 and 260 microbubbles/cm². Relative backscatter intensities were also drawn randomly, reflecting the backscatter intensity variations of a polydisperse microbubble population imaged at various distances from the elevational beam axis), and ranged between 0.4 and 1 (a.u.). The set of microbubble locations was then converted to radiofrequency CEUS signals using the modulated PSF, which were then envelope detected through the Hilbert transform, and subsequently down-sampled to an 8 times courser grid to yield the input patches. The local PSF was estimated by manually pinpointing several isolated microbubbles and fitting a 2D anisotropic rotated Gaussian to the data. Uncertainty in this estimate was incorporated in the training procedure by introducing variance in the PSF parameters $\varphi$ through a multiplicative random component, i.e. $\varphi = \varphi_m \cdot [1 + \mathcal{N}(\mu=0, \sigma=0.1)]$. To increase the trained model's robustness, we added white and colored background noise with relative standard deviations of 2% and 5%, respectively. Colored noise was produced by spatially filtering white noise with a 2D Gaussian having a standard deviation of 1.2 pixels.

**Deep neural network architecture**
We adopt a fully convolutional U-net style architecture[20] that consists of an encoder network which captures essential image information into a latent feature layer, and an expanding decoder network which maps this latent representation to precise localizations on a high-resolution grid. The encoder follows a contracting path which consists of 3 layer-blocks, each block comprising two 3x3 convolution layers with leaky rectified linear unit (ReLU) activations, and one 2x2 Max-pooling operation. We use leaky ReLUs[29] rather than regular ReLUs across all convolution layers in the network to avoid inactive neurons/nodes that effectively decrease the model capacity. In addition, batch normalization is used before all activations to boost the network's trainability by enabling higher learning rates and requiring less-strict hyper-parameter optimization[30].

The subsequent latent layer includes two 3x3 convolutional layers, followed by a dropout layer (probability 0.5) which randomly disables about 50% of the latent features during training. This latent space is then transformed to a high-resolution localization image by the decoder. The decoder again consists of 3 blocks; the first two blocks encompassing two 5x5 deconvolution layers (transposed convolution)[31] of which the second has an output stride of 2 rather than 1, followed by a 2x2 up-sampling layer which simply repeats the image rows and columns. The last block consists of two deconvolution layers, of which the second again has an output stride of 2, preceding another 5x5 convolution which maps the feature space to a single-channel image through a linear activation function. The full network effectively scales the input image dimensions up by a factor 8.

**Network training and cost function**
We used the Adam optimizer with learning rate 0.001, and trained the network across 20,000 iterations to minimize the following cost function, similar to the one proposed in[19]:

$$c(x, y | \theta) = \| f(x|\theta) - G * y \|_2^2 + \lambda \| f(x|\theta) \|_1, \tag{1}$$

where $x$ and $y$ are input CEUS and target super-resolved patches, respectively, $f(x|\theta)$ is the nonlinear neural network function with parameters (weights and biases) $\theta$, and $\lambda$ is a regularization parameter that promotes network predictions that yield sparse images, and was (conservatively) set equal to 0.01. The operator $G$ denotes a 2D Gaussian filter of which the standard deviation was set to one pixel. In practice, we observed that applying such a mild 2D filtering operation on the sparse target data improved training stability; small localization errors (e.g. one pixel) are penalized less than large errors. This mean-squared-error-based regression strategy enables joint estimation of microbubble locations and their backscatter intensities. The latter is particularly useful to emphasize localizations near the elevational beam axis during image reconstruction.

Training (and inference) were run on a computation server, equipped with an NVidia Titan X Pascal that has 12 GB of video memory.

**Standard ULM**

Standard ULM was implemented using a centroid localization approach, largely following the methodology described by Errico *et al*.[6] We first up-sample the images by a factor 8 (equal to the grid up-sampling of Deep-ULM) and deconvolve them with a Gaussian low-pass filter based on the PSF, keeping only values above 50% of the 98th percentile. We then perform a morphological opening operation to remove spurious peaks, after which we detect the local maxima. We finally select a small area of 24x24 pixels (i.e. 3x3 pixels on the original data) around the local maxima and therefrom compute the local image centroids. The code was written in Python, using the `scikit-image` and `scipy` modules.

**Sparse recovery based ULM**

Sparse recovery based ULM[14] approaches the microbubble localization task as an inverse problem, by modelling each image frame as a superposition of translated and scaled PSFs according to microbubble locations and backscatter amplitudes on a high-resolution grid. Assuming that the microbubbles are smaller than a pixel and sparsely distributed across the image, the following regularized inverse problem can be formulated by promoting a sparse solution through the addition of an $\ell_1$ penalty:

$$\hat{\mathbf{x}} = \underset{\mathbf{x}}{\mathrm{argmin}} \quad \|A\mathbf{x} - \mathbf{y}\|_2^2 + \lambda \|\mathbf{x}\|_1, \tag{2}$$

where **x** is the microbubble reflectivity vector on a high-resolution grid, **y** is the vectorized image frame, and A is the measurement matrix in which each column is a shifted version of the PSF. To solve (2), we employed a highly optimized Fourier domain implementation of the Fast Iterative Shrinkage-Thresholding Algorithm (FISTA). We used a grid up-sampling factor of 8, and $\lambda$ was set to 0.01.

**Sparse recovery via SUSHI**

Sparsity-based super-resolution hemodynamic imaging[15] aims at producing time-lapse super-resolved sequences of fast hemodynamic changes. It was implemented by first dividing the acquired CEUS clip into small movie segments and estimating the pixel-wise variance of each segment, resulting in a variance time-lapse movie of the MBs, which exhibits improved spatial resolution and background rejection. To further improve the spatial resolution beyond the acoustic diffraction limit, sparse recovery is then performed on the variance images, by using a similar formulation to (2), with an up-sampling factor of 8. In supplementary fig. 2, the two leftmost SUSHI recoveries were performed with $\lambda = 0.01$, the middle recovery with $\lambda = 0.005$, and the two rightmost recoveries were performed with $\lambda = 0.001$. The three rightmost recoveries were performed with weighted sparse recovery, in which the weighting vector is the inverse of the SUSHI recovered image with an up-sampling factor of 4, interpolated to the 8 times denser grid.

**Microbubble data association and tracking**

Following the detection of all individual microbubbles, an automatic data association and tracking algorithm was applied[8]. In particular, the multiple hypothesis tracking algorithm (MHT), as first suggested by Reid[32] is considered as one of the most popular automatic data association algorithms. Using the MHT algorithm, we accumulated detections of individual microbubbles and grouped them such that every group represents the spatial locations of each microbubble as time progresses. A new measurement includes the detected position of the MB using Deep-ULM and a crude velocity estimate by applying optical flow[33] to the acquired sequence. This procedure was also used in Triple-SAT[8] as means to improve the tracking performance. With every newly associated measurement, position and velocity are estimated using a Kalman filter with a simple linear propagation model. At the end of this process, maps of individual microbubble trajectories as well as their velocities are produced.

## ACKNOWLEDGEMENTS

The authors would like to thank the NVIDIA Corporation and its academic GPU program for donating a Titan X Pascal which greatly facilitated the research described in this work.

# AUTHOR CONTRIBUTIONS

R.v.S., H.W., M.M., M.B., Z.K. and Y.E. designed/performed the experiments; R.v.S. designed the algorithm; R.v.S., O.S., M.M. and Y.E. designed and performed the data analysis. All the authors discussed the results and wrote the paper.

# DATA AVAILIBILITY

The data that support the findings of this study are available from the corresponding author upon reasonable request.

# SUPPLEMENTARY DATA

Supplementary fig. 1 displays the recovered density and localization precision of Deep-ULM compared to standard ULM (deconvolution and centroid localization) and ULM based on sparse recovery[14] as a function of simulated microbubble density. A microbubble is only considered detected if a localization was obtained close to its true location, within 30 $\mu$m (about 1/7th of the wavelength). To determine the localization precision, each identified microbubble is associated to the closest ground-truth microbubble position, and their Euclidian distance is calculated. For low densities, all methods perform similarly well, with Deep-ULM displaying a very slight increase in localization error. When the density increases however, sparse recovery adequately detects more microbubbles than standard ULM, while Deep-ULM significantly outperforms both sparse recovery and standard ULM in terms of detection rate and localization precision.

A qualitative comparison of standard ULM, sparsity-driven ULM, SUSHI, and Deep-ULM is given in supplementary fig. 2. We simulated a 10-second ultrasound acquisition of microbubbles moving at 1 mm/s through 3 pairs of parallel vessels (separated by $\lambda/3$, $\lambda/4$, and $\lambda/5$, respectively) for increasing densities. From supplementary fig. 2C, we can observe that standard ULM again performs very well for low densities, but yields many false localizations when the number of microbubbles per area increases. SUSHI however displays good performance across all densities (supplementary fig. 2D); although it does not detect the most closely spaced vessels for the higher densities, it delineates the $\lambda/3$- and $\lambda/4$-separated vessels across all experiments. Sparsity-driven ULM (supplementary fig. 2E) remains more robust than standard ULM up to higher densities, but is less stable then SUSHI for the densities used in the two rightmost panels. Despite the use of a highly-optimized Fourier-domain implementation[15], the sparsity-based methods are about four orders of magnitude slower than inference with GPU-accelerated Deep-ULM (~6 seconds/frame compared to ~0.6 milliseconds/frame on our system). In this specific experiment, Deep-ULM outperforms the sparse recovery methods, which we attribute to learning of the image-domain interference patterns of closely-spaced microbubbles (supplementary figs. 2F-G).

Supplementary fig. 3 shows an illustrative example of Deep-ULM recovery corresponding to a single CEUS frame of a rat spinal cord recording. The method robustly resolves individual microbubbles on dense data with significant overlaps in their PSFs. Supplementary fig. 4 shows how Deep-ULM can be used in conjuncture with microbubble tracking to enable visualization of flow orientation.

In supplementary fig. 5, we demonstrate initial feasibility Deep-ULM in clinical practice, by performing inference on clinical CEUS measurements of a patient's prostate using a standard ultrasound system (Philips iU22) in combination with a standard transrectal probe. The frame rate of this clinical scanner was 10 Hz. We trained the neural network as before, requiring nothing more than an estimate of the PSF parameters. We then deploy it by probing the vasculature in an area located within the peripheral zone of the organ. Real-time inference rates above 20 Hz were achieved. The microbubble localizations shown in supplementary fig. 5B-C evidence how Deep-ULM adequately discerns overlapped responses. About 15,000 microbubbles were detected across 300 frames.

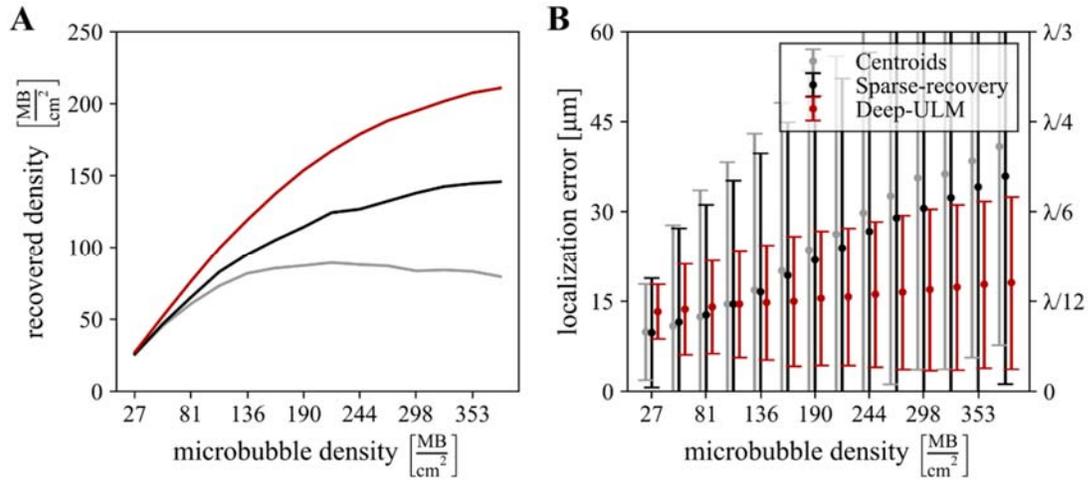

Supplementary figure 1. Detection rate and localization precision of Deep-ULM (red) compared to ULM based on centroids (gray) and sparse recovery (black). (**A**) Recovered density as a function of simulated microbubble (MB) density, and (**B**) corresponding median localization errors with bars representing the standard deviation. Note that Deep-ULM's localization errors are very close to the grid spacing (15 µm), and well below the wavelength (214 µm), even for high microbubble densities.

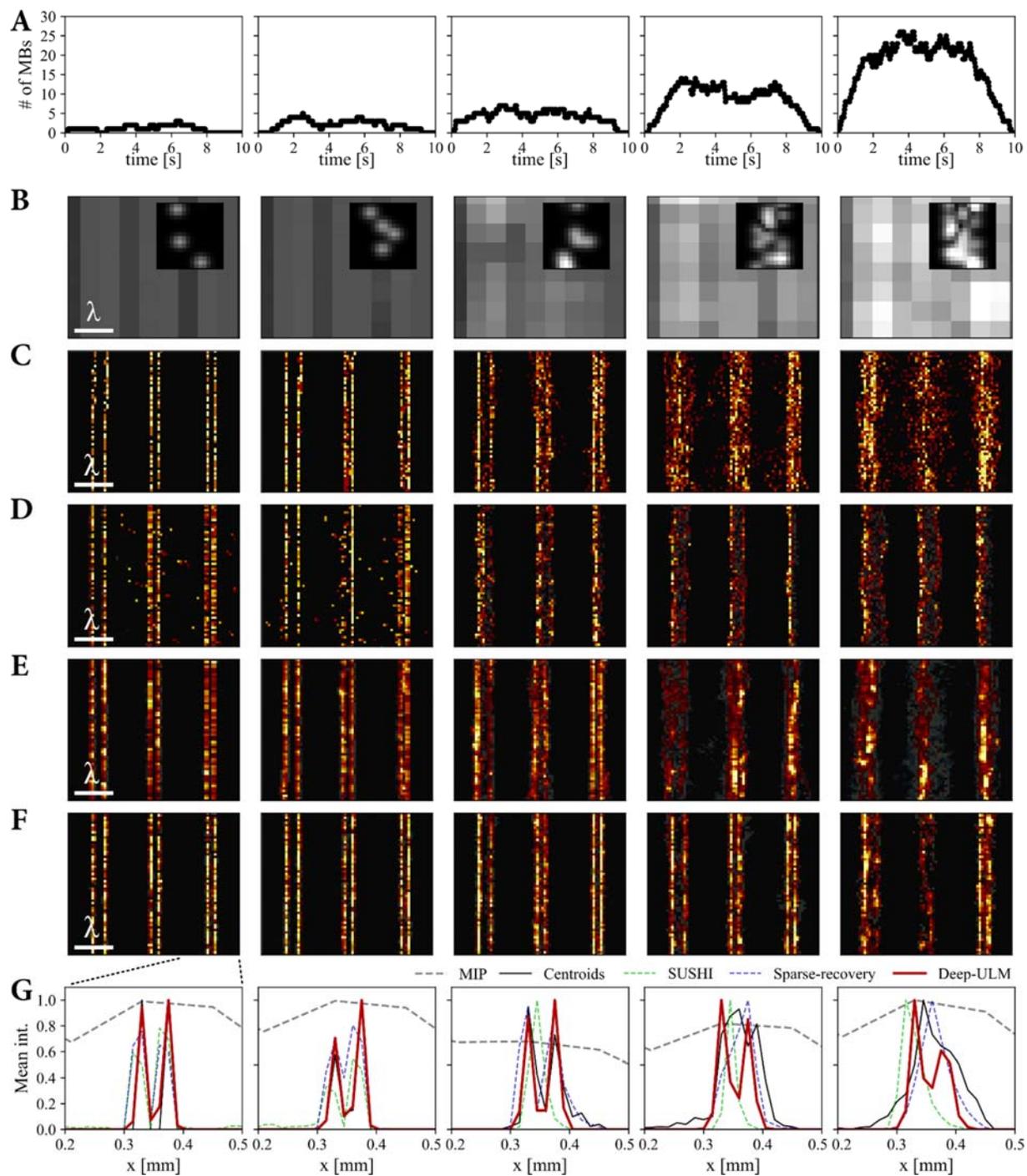

Supplementary figure 2. In-silico evaluation of Deep-ULM compared to ULM based on centroids and sparse-recovery for parallel microbubble streams with varying densities and interfering point spread functions. (**A**) Number of microbubbles in the frame across time, (**B**) Maximum intensity persistence images (MIP), along with individual example frames (inset), (**C**) standard centroid ULM images, (**D**) SUSHI, (**E**) sparse recovery ULM, (**F**) Deep-ULM, and (**G**) mean lateral profiles of the two rightmost streams for all techniques. Note that Deep-ULM attains better sub-wavelength separation for higher densities than the other methods.

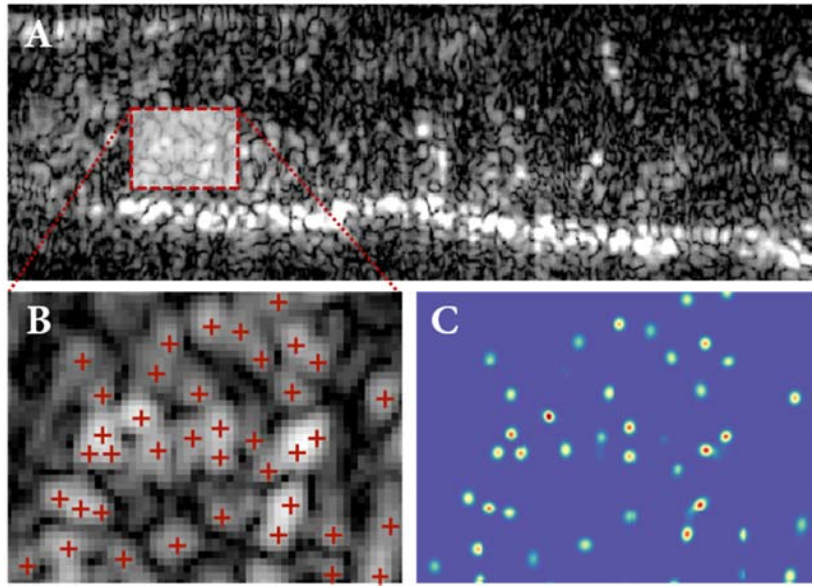

**Supplementary figure 3.** *In-vivo* **Deep-ULM localization examples on a rat spine.** (**A**) An illustrating example frame in the CEUS loop, (**B**) localizations within an area of interest of (**A**), and (**C**) corresponding Deep-ULM high-resolution recovery from which these localizations are deduced.

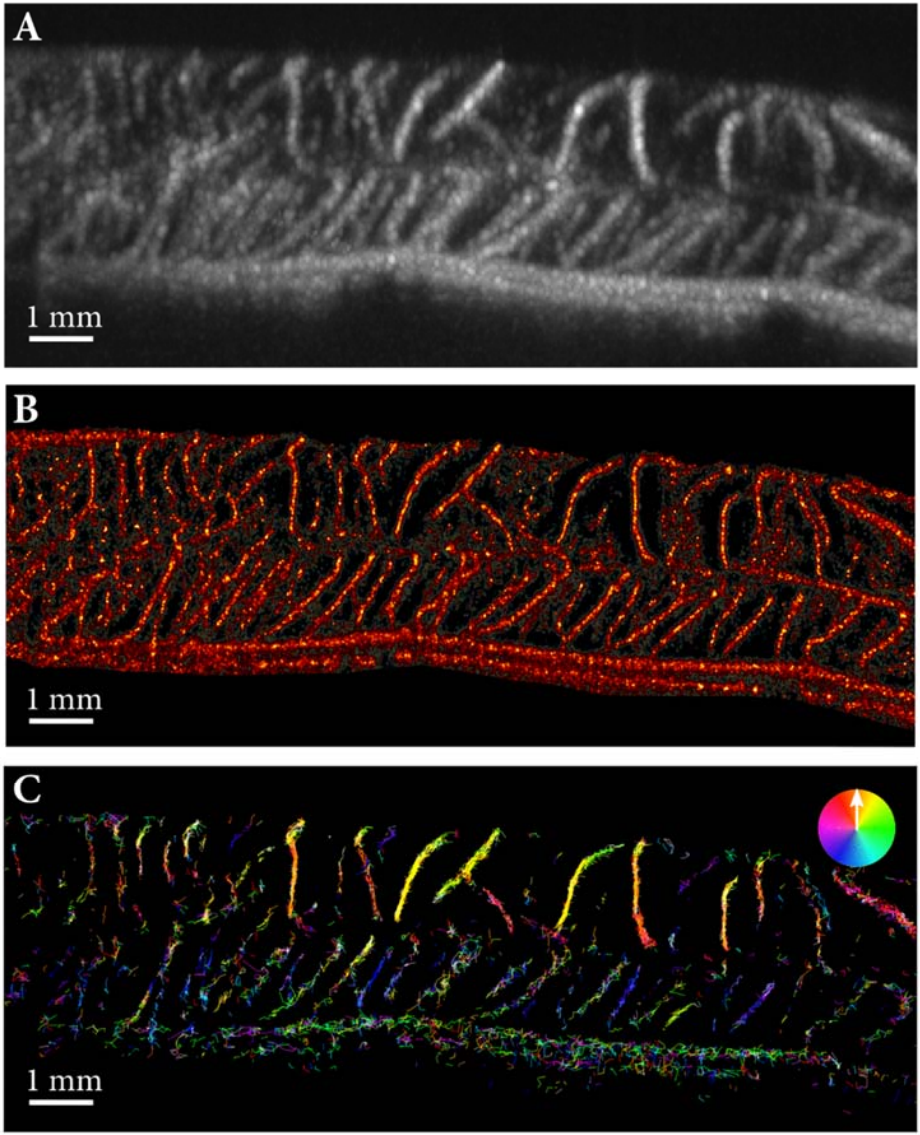

**Supplementary figure 4. Deep-ULM and tracking on a rat spine for a 2-second acquisition at a frame rate of 400 Hz. (A)** Maximum intensity persistence image, and **(B,C)** Deep-ULM image and microbubble tracking results, respectively.

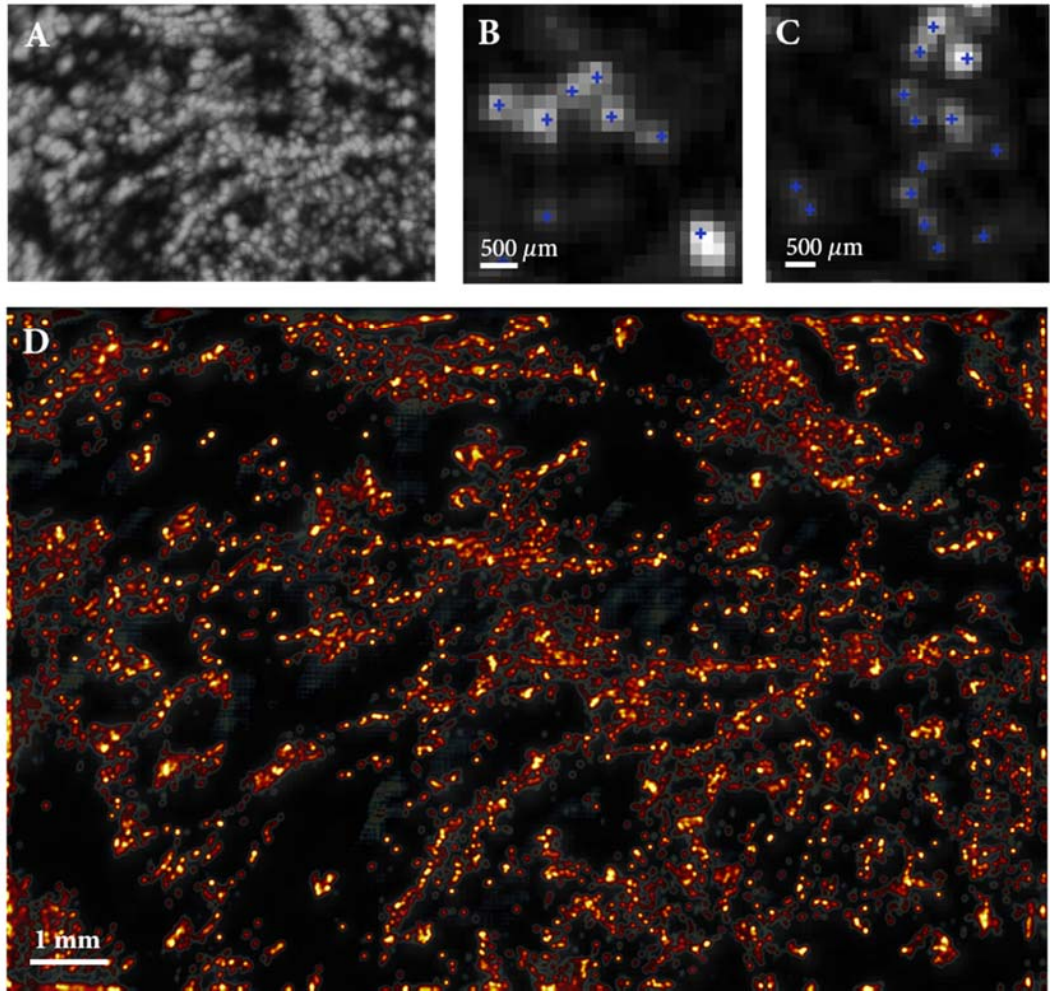

**Supplementary figure 5. Initial feasibility of** *in-vivo* **Deep-ULM on a human prostate.** (**A**) Maximum intensity persistence image across 300 imaging frames sampled at a frame rate of 10 Hz. (**B,C**) Examples of localizations in individual frames (**D**) Deep-ULM by summing reconstructed images across the same set of frames.